\documentclass[letterpaper,english,aps,prl,twocolumn,showpacs,10pt,superscriptaddress,floatfix,longbibliography]{revtex4-2}
\usepackage{mathptmx}
\usepackage[T1]{fontenc}
\usepackage[latin9]{inputenc}
\usepackage{color}
\usepackage{babel}
\usepackage{verbatim}
\usepackage{bm}
\usepackage{amsmath}
\usepackage{amssymb}
\usepackage{cancel}
\usepackage{graphicx}
\usepackage[normalem]{ulem}
\usepackage{soul}
\usepackage[unicode=true,pdfusetitle,
 bookmarks=true,bookmarksnumbered=false,bookmarksopen=false,
 breaklinks=false,pdfborder={0 0 1},backref=false,colorlinks=false]
 {hyperref}
\hypersetup{
 colorlinks=true,linkcolor=blue,anchorcolor=red,citecolor=blue, urlcolor=blue}

\makeatletter

\pdfpageheight\paperheight
\pdfpagewidth\paperwidth

\usepackage{mathptmx}
\usepackage{graphicx}
\usepackage{subfigure}
\usepackage{calrsfs}
\DeclareMathAlphabet{\pazocal}{OMS}{zplm}{m}{n}
\usepackage{amsmath}
\usepackage{amsfonts}
\usepackage{amssymb}
\usepackage{mathrsfs}
\usepackage{bm}
\usepackage{subfigure}
\usepackage{xcolor}
\usepackage{braket}
\usepackage{lineno}

\newlength{\seplinewidth}
\newlength{\seplinesep}
\colorlet{sepline}{orange}

\usepackage{pdfpages} 
\usepackage{pgffor} 
\usepackage{xr} 

\makeatletter
\AtBeginDocument{\let\LS@rot\@undefined}
\makeatother

\def\supplementfilename{supplement}

\pdfximage{\supplementfilename.pdf}
\def\numbersupplementpages{\the\pdflastximagepages}

\newif\ifarXiv
\arXivtrue 

\makeatother

\begin{document}
\title{Quantum Nonlinear Acoustic Hall Effect and Inverse Acoustic Faraday
Effect in Dirac Insulators}
\author{Ying Su}
\email{yingsu001@gmail.com}

\affiliation{Center for Integrated Nanotechnology, Los Alamos National Laboratory,
Los Alamos, New Mexico 87545, USA}
\author{Alexander V. Balatsky}
\affiliation{Nordita, KTH Royal Institute of Technology and Stockholm University
106 91 Stockholm, Sweden}
\affiliation{Department of Physics, University of Connecticut, Storrs, Connecticut
06269, USA}
\author{Shi-Zeng Lin}
\homepage{szl@lanl.gov}

\affiliation{Theoretical Division, T-4 and CNLS, Los Alamos National Laboratory,
Los Alamos, New Mexico 87545, USA}
\affiliation{Center for Integrated Nanotechnology, Los Alamos National Laboratory,
Los Alamos, New Mexico 87545, USA}
\begin{abstract}
{We propose 
to realize the quantum nonlinear Hall effect and the inverse Faraday effect through the acoustic wave in a time-reversal invariant but inversion broken Dirac insulator. We focus on the acoustic frequency much lower than the Dirac gap such that the interband transition is suppressed and these effects arise solely from the intrinsic valley-contrasting band topology.}
The corresponding acoustoelectric conductivity and magnetoacoustic
susceptibility are both proportional to the quantized valley
Chern number and independent of the quasiparticle lifetime. The linear and nonlinear components of the
longitudinal and transverse topological currents can be tuned
by adjusting the polarization and propagation directions of the surface
acoustic wave. The static magnetization generated by a
circularly polarized acoustic wave scales linearly with the acoustic
frequency as well as the strain-induced charge density. Our results
unveil a quantized nonlinear topological acoustoelectric response of gapped
Dirac materials, like hBN and transition-metal dichalcogenide,
paving the way toward room-temperature acoustoelectric devices due
to their large band gaps. 
\end{abstract}
\date{\today}

\maketitle


\emph{Introduction.}\textemdash Dirac materials, owing to their relativistic
energy dispersion and nontrivial topological properties, have drawn
significant attention \citep{Vafek2014Dirac,Wehling2014Dirac,Armitage2018Weyl}.
One important feature in many Dirac materials is the emergence of valley degree of freedom
around the low-energy band extrema where electronic states are well
separated in momentum space and form Kramers pairs in the presence
of time-reversal symmetry (TRS) \citep{rycerz2007valley,Xiao2007Valley}.
Similar to spin, the states with opposite valley indices can be used
as binary information carriers that evokes a surge of interest in
valleytronics \citep{schaibley2016valleytronics,vitale2018valleytronics}.
Recently, the acoustic wave has been demonstrated as an efficient
method to manipulate the valley degree of freedom and has led to
the valley acoustoelectric (AE) effect \citep{Kalameitsev2019Valley,Sonowal2020Acoustoelectric,Wan2024Strongly},
acoustogalvanic effect \citep{Sukhachov2020Acoustogalvanic,Zhao2022Acoustically,Bhalla2022Pseudogauge},
axial magnetoelectric effect \citep{Liang2021Axial}, etc. Within these effects,
 direct (valley) current and static magnetization are generated
through the nonlinear response to the alternating strain exerted by
different acoustic waves in doped Dirac semimetals and insulators \citep{Kalameitsev2019Valley,Sonowal2020Acoustoelectric,Wan2024Strongly,Sukhachov2020Acoustogalvanic,Zhao2022Acoustically,Bhalla2022Pseudogauge,Liang2021Axial}. 

The TRS-{invariant} strain field emerges
as a deformation potential as well as a valley-contrasting gauge field for Dirac fermions
\citep{Suzuura2002Phonons,Manes2007Symmetry,vozmediano2010gauge,guinea2010energy}, {that results in novel quantum effects like the zero-field Landau quantization \citep{guinea2010energy,levy2010strain}, quantum spin Hall effect \citep{Cazalilla2014Quantum}, valley polarization and inversion \citep{Settnes2016Graphene,Li2020Valley}, etc}.
Despite the conventional AE effect stemming from the deformation potential
had been identified in 1950s \citep{Parmenter1953AE,Weinreich1957AE},
the role of the emergent gauge field exerted by an acoustic wave passing
through the doped Dirac semimetal or insulator has just been recoginized
\citep{Sukhachov2020Acoustogalvanic,Zhao2022Acoustically,Bhalla2022Pseudogauge,Liang2021Axial}.
In particular, the interplay between the Dirac fermion and acoustic
wave gives rise to a direct Hall current transverse to the wave vector without 
the TRS breaking, similar to the nonlinear Hall effect
\citep{du2021nonlinear}.\textcolor{black}{{} To date, the AE Hall effect
arises from the second-order response to the emergent gauge field
of acoustic waves and exists only in the metallic phase} \citep{Sukhachov2020Acoustogalvanic,Zhao2022Acoustically,Bhalla2022Pseudogauge}.
Likewise, the inverse acoustic faraday effect (IAFE, which generates
static magnetization by a circularly polarized acoustic wave \citep{tokman2013inverse,shabala2024phonon})
can be achieved in Dirac semimetals via the second-order response that results in rectified magnetization
\citep{Liang2021Axial}.

Discovering the quantum version of various Hall effects is of significant importance in condensed matter physics by unveiling different
topological phases of insulators and quantized topological responses \citep{Klitzing1980New,Thouless1982Quantized,Haldane1988Model,Kane2005Quantum,Bernevig2006Quantum,konig2007quantum,chang2013experimental}.
However, it remains unclear how the undoped Dirac insulator (i.e.,
the Fermi energy lies in the band gap) responds to different acoustic
waves.  
In this work, we propose to realize the quantum
nonlinear acoustic Hall effect (QNAHE) and IAFE through the {quantized} topological
AE response in 2D Dirac insulators with TRS but no inversion symmetry.
In QNAHE, we show that a surface acoustic wave (SAW) passing through
the Dirac insulator can generate both longitudinal and transverse
currents. Moreover, their linear and nonlinear components (up to the
second order) are tunable by adjusting the polarization and propagation
directions of the SAW. When the acoustic wave is circularly polarized
and propagating in the out-of-plane direction, there is a static magnetization
in the Dirac insulator subject to inhomogeneous strain, i.e., the
realization of IAFE.
{Here we focus on the acoustic frequency much lower than the Dirac gap such that the interband transition is suppressed. Since no free charge carrier is involved in the AE response, the QNAHE and IAFE arise solely from the intrinsic valley-contrasting band topology of the Dirac insulator.}
In both cases, the corresponding AE conductivity
and magnetoacoustic (MA) susceptibility are proportional to the quantized
valley Chern number and independent of the quasiparticle lifetime, {manifesting their topological origins}. Our results provide a new mechanism of rectification
that converts the alternating acoustic wave into direct (Hall) current
and static magnetization in the Dirac insulator via the 
{quantized} nonlinear
topological AE response.

\begin{figure}
\begin{centering}
\includegraphics[width=8cm]{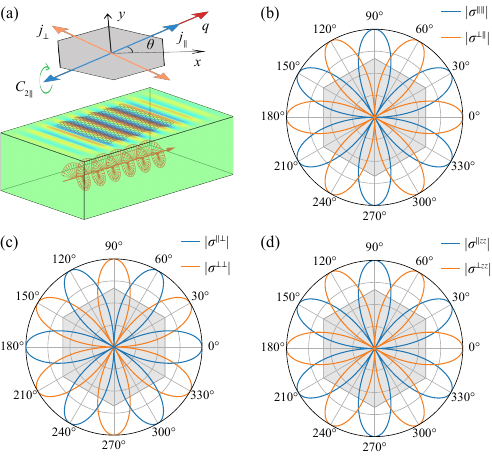}
\par\end{centering}
\caption{(a) 
Schematic setup for the QNAHE. The upper panel shows the directions
of the SAW vector $\bm{q}$, longitudinal current $j_{\parallel}$,
and transverse current $j_{\perp}$ with respect to the $x$ and $y$
axes that are fixed along the zigzag and armchair directions of the
honeycomb lattice, respectively. $C_{2\parallel}$ represents the two-fold rotation
along the axis of the SAW wavevector $\bm{q}$. The lower panel depicts
a 2D Dirac material placed on a piezoelectric substrate in which a SAW propagates
along its wave vector direction. (b) and (c) The polar plots of the absolute value of the
linear AE conductivities. (d) The polar plot of the absolute value of the nonlinear
AE conductivities. \label{fig:fig1}}
\end{figure}

\emph{Model and topological acoustoelectric response.}\textemdash Here
we consider the low-energy states of a 2D Dirac insulator, like hBN
or transition-metal dichalcogenide, which can be generally described
by the Dirac action
\begin{equation}
S=\int d^{3}x\bar{\psi}\left(i\cancel{\partial}+\cancel{A}+\tau_{z}\cancel{a}-\tau_{z}m\right)\psi,\label{eq:S0}
\end{equation}
where $\cancel{\partial}=\gamma^{\mu}\partial_{\mu}$ denotes the
Feynman slash notation and $\gamma^{\mu=0,1,2}=\left(\sigma_{z},i\sigma_{y},-i\sigma_{x}\right)$
with $\sigma_{i}$ being the Pauli matrix for internal degrees of
freedom, e.g., sublattice or orbital. $\bar{\psi}=\psi^{\dagger}\gamma^{0}$
represents the Dirac adjoint. For convenience, we set the units $e=\hslash=v_{F}=1$
where $v_{F}$ is the {Dirac} velocity. $A_{\mu}=\left(\Phi,-\bm{A}\right)$
is the real gauge field, while $a_{\mu}=\left(0,-\bm{a}\right)$ is
the emergent gauge field. In the presence of a strain field $\bm{u}(\bm{r})$,
the deformation potential is $\Phi=g_{D}(u_{xx}+u_{yy})$ and the
emergent gauge field reads $\bm{a}=g\left(u_{yy}-u_{xx},2u_{xy}\right)$
where $u_{ij}=\frac{1}{2}\left(\partial_{i}u_{j}+\partial_{j}u_{i}+\partial_{i}d\partial_{j}d\right)$
is the strain tensor \citep{Suzuura2002Phonons,Manes2007Symmetry,vozmediano2010gauge}.
Here $d(\bm{r})$ is the out-of-plane displacement field and $g=\beta/2a_{0}$
with the Gr{\"u}neisen parameter $\beta$ and lattice bond length
$a_{0}$. Due to the TRS, the emergent gauge field and Dirac mass
are valley-contrasting, as described by the last two terms of Eq.~(\ref{eq:S0}) where $\tau_{i}$ is the Pauli matrix for valley degrees
of freedom. 

To investigate how the Dirac insulator responds to the emergent gauge
field, we integrate out the fermion in Eq.~(\ref{eq:S0}) which yields
a Chern-Simons (CS) action \citep{Vaezi2013Topological}
\begin{equation}
S_{\mathrm{CS}}=\frac{\mathrm{sgn}(m)}{2\pi}\int d^{3}x\epsilon^{\mu\nu\lambda}A_{\mu}\partial_{\nu}a_{\lambda},\label{eq:SCS}
\end{equation}
where $\mathrm{sgn}(m)=C_{K}-C_{K^{\prime}}$ is the valley Chern
number. In contrast to the conventional CS term that vanishes due
to TRS, Eq.~(\ref{eq:SCS}) represents a cross CS term between the real
and emergent gauge fields, leading to the topological current density 
\begin{equation}
j^{\mu}=\frac{\delta S_{\mathrm{CS}}}{\delta A_{\mu}}=\frac{\text{sgn}(m)}{\pi}\epsilon^{\mu\nu\lambda}\partial_{\nu}a_{\lambda}.\label{eq:ju}
\end{equation}
In terms of the emergent gauge field, the pseudo-electric and magnetic
fields are defined as $\bm{E}_{s}=-\partial_{t}\bm{a}$ and $B_{s}=\left(\nabla\times\bm{a}\right)\cdot\hat{\bm{z}}$.
Then Eq.~(\ref{eq:ju}) can be re-expressed as
\begin{equation}
\bm{j}=\frac{2e^{2}\mathrm{sgn}(m)}{h}\left(\begin{array}{cc}
0 & -1\\
1 & 0
\end{array}\right)\bm{E}_{s},\quad\rho=-\frac{2e^{2}\mathrm{sgn}(m)}{h}B_{s},\label{eq:jEs}
\end{equation}
where the physical units are recovered. It demonstrates that: (i)
there is a current transverse to the pseudo-electric field; (ii) The
charge density is proportional to the pseudo-magnetic field; (iii)
Both current and charge densities are proportional to the quantized
valley Chern number. 

In principal, Eqs.~(\ref{eq:ju}) and (\ref{eq:jEs}) reflect the
linear response to the emergent gauge field and hence the alternating
in-plane strain generates only the alternating linear current
in graphene \citep{Vaezi2013Topological,Sela2020Quantum}. Meanwhile,
the strain-induce charge density yields the piezoelectricity \citep{Droth2016Piezoelectricity,rostami2018piezoelectricity}
and magnetoelectric effect \citep{Rodriguez2018Theory} in the Dirac
system. Nevertheless, there exists a second-order response to the strain in Eq.~(\ref{eq:ju})
since the emergent gauge field is proportional to the strain tensor
that is quadratic in the gradients of the out-of-plane displacement
$d(\bm{r})$. To the best of our knowledge, this nonlinear topological
AE response stemming from the CS term and acoustic wave in the Dirac
insulator has not been discussed before and is the main focus of this
study {by giving rise to the QNAHE and IAFE. Using the realistic material parameters, we will show that the acoustic-wave-induced direct topological current and static magnetization in Dirac insulators are measurable with the magnitudes comparable to other nonlinear responses. } 

\emph{Quantum nonlinear acoustic Hall effect.}\textemdash To trigger
the topological current in Eq.~(\ref{eq:jEs}), we consider a generic
SAW 
\begin{equation}
\bm{u}=\mathrm{Re}\left[\left(u_{\parallel}\hat{\bm{q}}+u_{\perp}\partial_{\theta}\hat{\bm{q}}-iu_{z}\hat{\bm{z}}\right)e^{i(\bm{q}\cdot\bm{r}-\omega t)}\right],\label{eq:u}
\end{equation}
passing through the Dirac insulator, as shown in Fig.~\ref{fig:fig1}(a).
Here the 2D material is placed on top of a piezoelectric substrate
and the SAW can be launched by the interdigital transducer \citep{white1965direct}.
The wave vector $\bm{q}=q\left(\cos\theta,\sin\theta\right)$ is measured
with respect to the $x$ and \emph{$y$} axes that are fixed along
the zigzag and armchair directions of the honeycomb lattice respectively, see Fig.~\ref{fig:fig1}(a). When $u_{\parallel}=0$ ($u_{\perp}=0$), Eq.~(\ref{eq:u}) describes the circularly polarized (Rayleigh) SAW. When
$u_{\parallel}=u_{z}=0,$ it becomes the Bleustein-Gulyaev SAW \citep{biryukov1995surface}.
The topological current driven by the SAW can be decomposed into the
longitudinal and transverse components as $\bm{j}=j_{\parallel}\hat{\bm{q}}+j_{\perp}\partial_{\theta}\hat{\bm{q}}$
with $j_{\alpha}=\sum_{n}\mathrm{Re}[j_{\alpha}^{(n)}e^{i(n\bm{q}\cdot\bm{r}-n\omega t)}]$
whose nonzero components are
\begin{align}
\left(\begin{array}{c}
j_{\parallel}^{(1)}\\
j_{\perp}^{(1)}
\end{array}\right)= & \frac{2e^{2}\mathrm{sgn}(m)gq\omega}{h}\left(\begin{array}{c}
u_{\parallel}\sin3\theta+u_{\perp}\cos3\theta\\
u_{\parallel}\cos3\theta-u_{\perp}\sin3\theta
\end{array}\right),\label{eq:j1}
\end{align}
\begin{equation}
\left(\begin{array}{c}
j_{\parallel}^{(2)}\\
j_{\perp}^{(2)}
\end{array}\right)=-i\frac{e^{2}\mathrm{sgn}(m)gq^{2}\omega u_{z}^{2}}{h}\left(\begin{array}{c}
\sin3\theta\\
\cos3\theta
\end{array}\right).\label{eq:j2}
\end{equation}
A nonlinear (second-order) topological current can be
generated by the out-of-plane component $u_{z}$ of the SAW, besides
a linear (first-order) topological current from the in-plane components
$u_{\parallel}$ and $u_{\perp}$. 

Now we introduce the AE conductivity tensors for linear and nonlinear
currents as
\begin{equation}
j_{\alpha}^{(1)}=\sigma^{\alpha\beta}u_{\beta},\qquad\qquad j_{\alpha}^{(2)}=\sigma^{\alpha\beta\gamma}u_{\beta}u_{\gamma}.\label{eq:AEC}
\end{equation}
According to Eqs.~(\ref{eq:j1}) and (\ref{eq:j2}), the tensors are
proportional to the acoustic frequency as well as the quantized valley
Chern number {and are independent of the quasiparticle lifetime}. Furthermore, their polar angle dependence exhibits the
$C_{3z}$ symmetry, consistent with the lattice rotational symmetry.
In Figs.~\ref{fig:fig1}(b) and \ref{fig:fig1}(c), we show the polar
plots of the absolute rank-2 AE conductivity tensors $|\sigma^{\parallel\alpha}|$
and $|\sigma^{\perp\alpha}|$ (for $\alpha=\parallel$ or $\perp$)
that are responsible for the linear longitudinal current $j_{\parallel}^{(1)}$
and transverse current $j_{\perp}^{(1)}$ driven by $u_{\alpha}$.
Note that the linear currents driven by $u_{\parallel}$ and $u_{\perp}$
are always perpendicular to each other, as elucidated in Eq.~(\ref{eq:j1}).
For the absolute rank-3 conductivity tensors $|\sigma^{\parallel zz}|$
and $|\sigma^{\perp zz}|$ in Fig.~\ref{fig:fig1}(d), they display
the same polar angle dependence as $|\sigma^{\parallel\parallel}|$
and $|\sigma^{\perp\parallel}|$ in Fig.~\ref{fig:fig1}(b). Thus
the nonlinear current driven by $u_{z}$ is always parallel (perpendicular)
to the linear current driven by $u_{\parallel}$ ($u_{\perp}$). In
particular for the SAW propagating in the zigzag (armchair) direction {with $\theta=n\pi/3$ ($n\pi/3+\pi/6$)},
$\sigma^{\parallel\parallel}=\sigma^{\perp\perp}=\sigma^{\parallel zz}=0$
($\sigma^{\perp\parallel}=\sigma^{\parallel\perp}=\sigma^{\perp zz}=0$).
This is constrained by the symmetry of the system. Under the two-fold
rotation $C_{2\parallel}$ along the axis of the SAW vector $\bm{q}$,
as shown in Fig.~\ref{fig:fig1}(a), $u_{\parallel}$ and $j_{\parallel}$
($u_{\perp}$, $u_{z}$ and $j_{\perp}$) are even (odd). When the
SAW propagates in the zigzag (armchair) direction, the two sublattices are exchanged (invariant) under the $C_{2\parallel}$ such that the Dirac mass $m$ is odd (even). 
Because the two sides of Eqs.~(\ref{eq:j1}) and (\ref{eq:j2})
have to transform in the same manner under the symmetry operations like $C_{2\parallel}$ and TRS,
we can explain the behaviors of the AE conductivities above, see the Supplemental Material (SM) \citep{SM} for detailed symmetry analysis. 

Exploiting the anisotropy of AE conductivity tensors, the linear and
nonlinear components of longitudinal and transverse topological currents
can be feasibly tuned by adjusting the polarization and propagation
directions of the SAW. For instance, the Rayleigh (circularly polarized)
SAW generates parallel (perpendicular) linear and nonlinear currents.
When the circularly polarized SAW propagates in the zigzag (armchair)
direction, there are linear (nonlinear) longitudinal current and nonlinear
(linear) transverse current. In practice, as long as the SAW has an
out-of-plane component and does not propagate in the armchair direction,
$\sigma^{\perp zz}\neq0$ leads to a nonlinear transverse current,
i.e., the QNAHE. 

The 
{quantized nonlinear topological AE} response originating from the quadratic term in the
strain tensor enables not only the frequency multiplier but also the
rectification. To realize the rectified direct topological current,
we consider a pulsed SAW whose amplitude modulation can be encoded
in the time-dependent $u_{\alpha}(t)$. Then the zeroth-order topological
current reads
\begin{equation}
\left(\begin{array}{c}
j_{\parallel}^{(0)}\\
j_{\perp}^{(0)}
\end{array}\right)=\frac{e^{2}\mathrm{sgn}(m)gq^{2}u_{z}\partial_{t}u_{z}}{h}\left(\begin{array}{c}
\sin3\theta\\
\cos3\theta
\end{array}\right).\label{eq:j0}
\end{equation}
By demanding $\partial_{t}\left(u_{z}\partial_{t}u_{z}\right)=0$, we
find that the sawtooth-type SAW with $u_{z}(t)=\bar{u}_{z}\sqrt{(t\mathrm{\;mod}\;T)/T}$
(where $T$ is the period of the pulse) can generate a direct topological current 
\begin{equation}
\bm{j}_{\mathrm{DC}}^{(0)}=\frac{e^{2}\mathrm{sgn}(m)gq^{2}\bar{u}_{z}^{2}}{2hT}(\sin3\theta,\cos3\theta),\label{eq:jDC}
\end{equation}
within each period and propagating in the same direction as the nonlinear
current in Eq.~(\ref{eq:j2}). To estimate the direct current density,
we take monolayer hBN with the Gr{\"u}neisen parameter $\beta=3.3$
and bond length $a_{0}=1.44$ $\textup{\AA}$ \citep{Droth2016Piezoelectricity}
as an example. For a typical pulsed SAW with $\bar{u}_{z}\sim10$
nm, $q\sim10$ $\mu\mathrm{m}^{-1}$, and $T\sim1$ $\mu\mathrm{s}$,
the current density is $j_{\mathrm{DC}}^{(0)}\sim15$ nA/cm, comparable
to that in the doped Dirac insulator \citep{Kalameitsev2019Valley}
and semimetal \citep{Sukhachov2020Acoustogalvanic}. 

Besides the topological current, the emergent gauge field generates
a charge density that satisfies the continuity equation $\partial_{\mu}j^{\mu}=0$,
as demonstrated in Eq.~(\ref{eq:ju}). For the SAW in Eq.~(\ref{eq:u}),
the charge density is $\rho=\sum_{n}\mathrm{Re}[\rho^{(n)}e^{i(n\bm{q}\cdot\bm{r}-n\omega t)}]$
with nonzero components $\rho^{(1)}=2e^{2}h^{-1}\mathrm{sgn}(m)gq^{2}(u_{\parallel}\sin3\theta+u_{\perp}\cos3\theta)$
and $\rho^{(2)}=-ie^{2}h^{-1}\mathrm{sgn}(m)gu_{z}^{2}q^{3}\sin3\theta$
according to Eq.~(\ref{eq:jEs}). To verify the topological AE response,
we also calculate the charge density from the tight-binding model
$H=\sum_{i}\varepsilon_{i}c_{i}^{\dagger}c_{i}+\sum_{\left\langle ij\right\rangle }t_{ij}c_{i}^{\dagger}c_{j}$
where $\varepsilon_{i}=\pm m$ is the sublattice potential and $t_{ij}=t_{0}\exp\left[-\beta\left(d_{ij}/a_{0}-1\right)\right]$
is the hopping strength in the honeycomb lattice. Here the bond length
$d_{ij}$ is altered by the SAW and the low-energy states of the lattice
model is described by Eq.~(\ref{eq:S0}). Using the realistic parameters
of the monolayer hBN, the charge densities from the CS theory and tight-binding
model exhibit excellent agreement with each other, see SM \citep{SM}
for details.

\begin{figure}
\begin{centering}
\includegraphics[width=8.5cm]{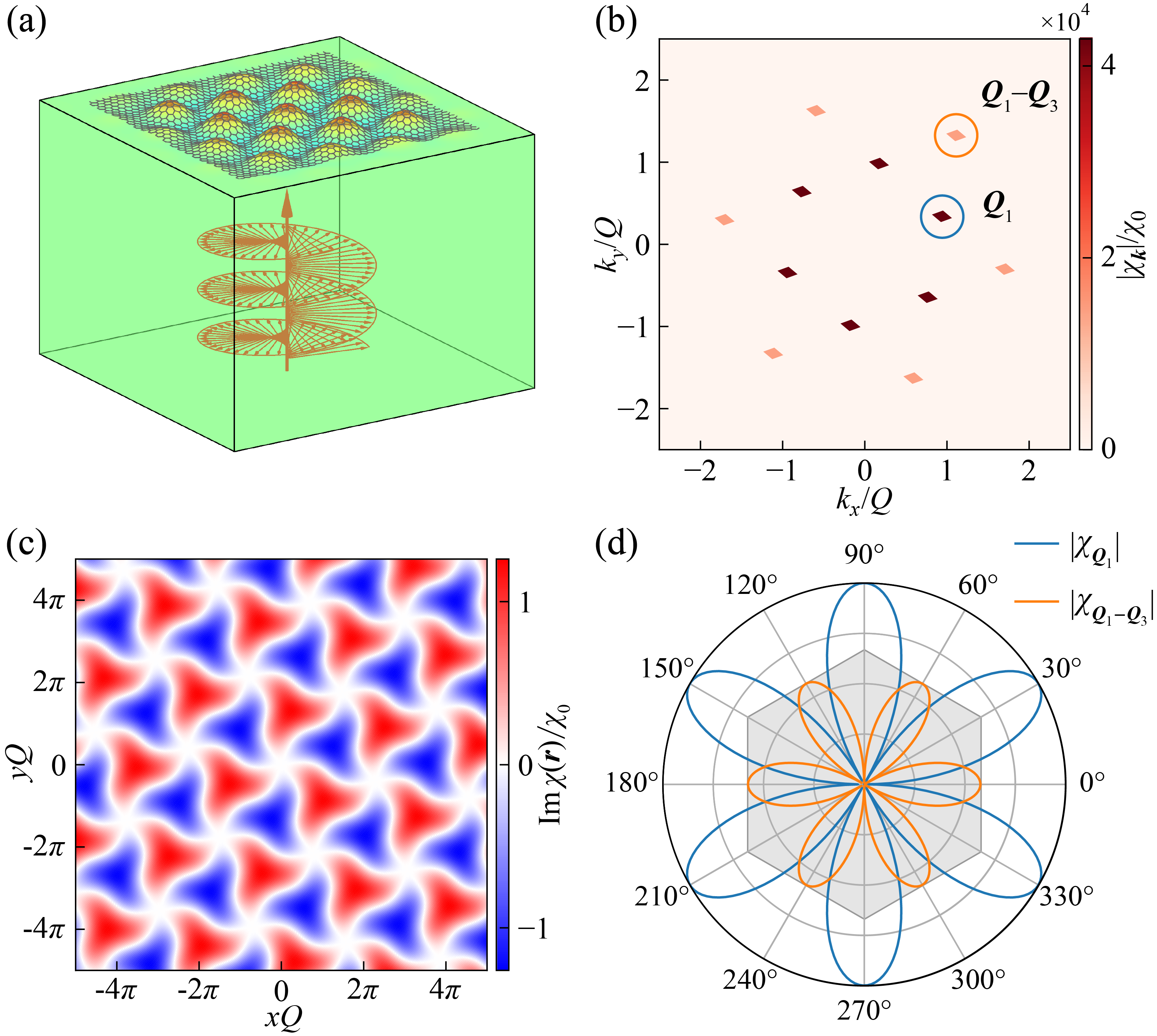}
\par\end{centering}
\caption{(a) Schematic setup for the IAFE. A periodically rippled 2D Dirac material
is placed on a piezoelectric substrate in which a circularly polarized
acoustic wave propagates in the out-of-plane direction. (b) and (c)
The renormalized MA susceptibilities in momentum and real space, respectively.
The blue and orange circles in (b) enclose the peaks at $\bm{Q}_{1}$
and $\bm{Q}_{1}-\bm{Q}_{3}$. (d) The polar plot of the MA susceptibility
peaks $\left|\chi_{\bm{Q}_{1}}\right|$ and $\left|\chi_{\bm{Q}_{1}-\bm{Q}_{3}}\right|$.
\label{fig:fig2}}
\end{figure}

\emph{Inverse acoustic Faraday effect.}\textemdash The charge density
induced by inhomogeneous strain through the topological AE response
generates not only piezoelectricity \citep{Droth2016Piezoelectricity,rostami2018piezoelectricity,Rodriguez2018Theory}
but also MA susceptibility. We now evaluate the static magnetization
excited by a circularly polarized acoustic wave in a strained Dirac
insulator, i.e., the realization of IAFE. As displayed in Fig.~\ref{fig:fig2}(a),
we consider a rippled 2D material on top of a non-flat piezoelectric
substrate and the inhomogeneous strain can be determined by minimizing
the elastic energy functional
\begin{equation}
V_{\mathrm{E}}=\int d^{2}\bm{r}\left[\frac{\lambda}{2}\left(\sum_{i}u_{ii}\right)^{2}+\mu\sum_{i,j}u_{ij}^{2}\right]+\frac{\kappa}{2}\int d^{2}\bm{r}\left(\nabla^{2}d\right)^{2},
\end{equation}
where $\lambda$ and $\mu$ are the Lam{\'e} factors and $\kappa$
is the bending rigidity \citep{Suzuura2002Phonons,Guinea2008Gauge}.
The topography of the 2D material is encoded in $d(\bm{r})$. Then
the electric charge due to strain-induced emergent gauge field
can be driven into orbital motion by a circularly polarized acoustic
wave $\bm{u}^{\prime}=\mathrm{Re}[u^{\prime}(\hat{\bm{x}}-i\hat{\bm{y}})e^{i(q_{z}z-\omega t)}]$
propagating in the out-of-plane direction. This yields a static magnetization 
\begin{equation}
\bm{M}_{\bm{k}}=\chi_{\bm{k}}\bm{u}_{q_{z},\omega}^{\prime}\times\bm{u}_{q_{z},\omega}^{\prime*},\label{eq:Mk}
\end{equation}
where $\bm{u}_{q_{z},\omega}^{\prime}=\left(u^{\prime}/2,-iu^{\prime}/2\right)$
and the MA susceptibility reads 
\begin{equation}
\begin{split}
\chi_{\bm{k}}= & \frac{2e^{2}\mathrm{sgn}(m)g\omega(\lambda+\mu)}{h(\lambda+2\mu)}\\
 & \times\frac{k_{y}\left(3k_{x}^{2}-k_{y}^{2}\right)\left[k_{x}^{2}f_{\bm{k},yy}-2k_{x}k_{y}f_{\bm{k},xy}+k_{y}^{2}f_{\bm{k},xx}\right]}{\left(k_{x}^{2}+k_{y}^{2}\right)^{2}}.\label{eq:chik}
\end{split}
\end{equation}
Here $\bm{u}^{\prime}(\bm{r})=\sum_{\bm{k}}\bm{u}_{\bm{k},\omega}^{\prime}e^{i(\bm{k}\cdot\bm{r}-\omega t)}$,
$d(\bm{r})=\sum_{\bm{k}}d_{\bm{k}}e^{i\bm{k}\cdot\bm{r}}$, and $\partial_{i}d\partial_{j}d=\sum_{\bm{k}}f_{\bm{k},ij}e^{i\bm{k}\cdot\bm{r}}$
with $f_{\bm{k},ij}=-\sum_{\bm{k}'}k_{i}^{\prime}\left(k_{j}-k_{j}^{\prime}\right)d_{\bm{k}^{\prime}}d_{\bm{k}-\bm{k}^{\prime}}$,
see SM \citep{SM} for details. As demonstrated in Eq.~(\ref{eq:chik}),
the MA susceptibility is proportional to the acoustic frequency as
well as the quantized valley Chern number {and is independent of the quasiparticle lifetime}, reflecting its topological origin. {This behavior is in contrast to the conventional inverse Faraday effect \citep{pitaevskii1961electric,Ziel1965Optically,Pershan1966Theoretical}, where the magneto-optic susceptibility is inversely proportional to the optical frequency in the dirty limit \citep{Battiato2014Quantum}.}
Meanwhile, the magnetization is proportional to the {strain-induced} charge density
and hence pseudo-magnetic field in Eq.~(\ref{eq:jEs}), which indicates a conversion of the emergent field into physical {magnetization}.

To be concrete, we consider a periodically rippled Dirac material
with the topography depicted by $d(\bm{r})=d_{0}\sum_{j=1}^{3}\cos\left(\bm{Q}_{j}\cdot r\right)$
\citep{Vazquez2008Periodically,guinea2009gauge} where $\bm{Q}_{j}=R_{3}^{j-1}\bm{Q}$
with $\bm{Q}=Q(\cos\theta,\sin\theta)$ and the three-fold rotation
operator $R_{3}$. For $\theta=20^{\circ}$, the renormalized MA susceptibility
$\chi_{\bm{k}}/\chi_{0}$ is shown in Fig.~\ref{fig:fig2}(b), where
the factor
\begin{equation}
\chi_{0}=\frac{2e^{2}\mathrm{sgn}(m)g\omega(\lambda+\mu)Q^{3}d_{0}^{2}}{h(\lambda+2\mu)},\label{eq:chi0}
\end{equation}
characterizes the strength of the MA susceptibility. Interestingly,
there are twelve peaks in Fig.~\ref{fig:fig2}(b), including six at
the wave vectors $\pm\bm{Q}_{j}$ of $d(\bm{r})$ and the others at
$\pm(\bm{Q}_{j+1}-\bm{Q}_{j})$ where $\bm{Q}_{4}\equiv\bm{Q}_{1}$. It indicates that $\chi(\bm{r})=\sum_{\bm{k}}\chi_{\bm{k}}e^{i\bm{k}\cdot\bm{r}}$
contains components whose period are smaller than that of $d(\bm{r})$, {resulting from the nonlinear terms in Eq.~(\ref{eq:chik})}.
In Fig.~\ref{fig:fig2}(c), we show $\mathrm{Im}\chi(\bm{r})/\chi_{0}$
that exhibits the antiferromagnetic signature. Note that $\chi(\bm{r})$
is purely imaginary such that the magnetization $\bm{M}(\bm{r})=\chi(\bm{r})\bm{u}_{q_{z},\omega}^{\prime}\times\bm{u}_{q_{z},\omega}^{\prime*}$
is real. Moreover, the ratio between $|\chi_{\bm{Q}_{1}}|$ and $|\chi_{\bm{Q}_{1}-\bm{Q}_{3}}|$
can be controlled by the polar angle $\theta$, as shown in Fig.~\ref{fig:fig2}(d).
In particular for $\bm{Q}_{j}$ along the zigzag (armchair) direction,
$|\chi_{\bm{Q}_{1}}|$ ($|\chi_{\bm{Q}_{1}-\bm{Q}_{3}}|$) vanishes,
leading to the magnetization with different periods, see SM \citep{SM}
for details. For $\omega\sim2\pi\times1$ GHz, $Q\sim1$ $\mathrm{nm^{-1}}$,
$d_{0}\sim1$ nm, $\lambda=3.5$ $\mathrm{eV}/\textup{\AA}^{2}$, and
$\mu=7.8$ $\mathrm{eV}/\textup{\AA}^{2}$ of monolayer hBN \citep{jung2015origin},
we estimate $\chi_{0}\sim0.4$ $\mu_{N}/\mathrm{nm^{4}}$ where $\mu_{N}$
is the nuclear magneton. For $u^{\prime}\sim10$ nm, it corresponds
to a magnetic momentum $m_{z}\sim1.2$ $\mu_{N}$ per unit cell, comparable
to the orbital magnetic moments of phonons in ionic materials \citep{Juraschek2017Dynamical,Juraschek2019Orbital}. 

The ferromagnetism can also be realized by the IAFE, which requires
the charge density induced by pseudo-magnetic field to be uniform in space.
Here we consider a triaxial stretch of the honeycomb lattice \citep{guinea2010energy,Neek2013Electronic,Settnes2016Pseudomagnetic}
with $(u_{r},u_{\theta})=u_{0}^{-1}r^{2}(\cos(3\theta-3\phi),\sin(3\theta+3\phi))$
in the polar coordinate, which leads to a uniform $\rho=-16e^{2}h^{-1}\mathrm{sgn(m)}gu_{0}^{-1}\sin3\phi$
in the bulk. Here $\phi$ is the polar angle of the stretch direction.
The static magnetization generated by the circularly polarized acoustic
wave is $\bm{M}=\frac{1}{2}\rho\bm{u}^{\prime}\times\dot{\bm{u}}^{\prime}=-\frac{1}{2}\omega\rho u^{\prime2}\hat{\bm{z}}.$
For $\theta=90^{\circ}$and $u_{0}\sim100$ nm, the estimated magnetization
is $M\sim3$ $\mu_{N}/\mathrm{nm^{2}}$ for the same circularly polarized
acoustic wave above. Note that the whole system remains charge neutral
at half filling in the gapped phase. Therefore, the net
charge in the bulk must be canceled by the opposite charge at boundaries,
see SM \citep{SM} for details. 

\textit{Discussion and conclusion.}\textemdash Here we compare the
proposed QNAHE and IAFE with the existing nonlinear Hall effect and
inverse Faraday effect. Despite the fundamental difference between
the acoustic wave and electromagnetic field as driving forces of
both effects, we further distinguish them from multiple perspectives.
So far, there are several different mechanisms of the nonlinear Hall
effect \citep{du2021nonlinear} that include the Berry curvature dipole
\citep{Sodemann2015Quantum}, quantum metric \citep{Gao2014Field},
disorder scattering \citep{du2019disorder}, and virtual interband
transition \citep{kaplan2023general}. {Among them, the first three
depend on the geometric and topological properties of the electron wave function at the Fermi energy and apply
only to the metal}, while the last one works for the magnetic insulator
without TRS and inversion symmetry. In contrast, QNAHE exists in the
Dirac insulator with TRS but with no inversion symmetry. {The disorder scattering
and virtual interband transition, which can be substantially suppressed by
the large band gap, are irrelevant in QNAHE}. Compared to
the quantum valley Hall effect of the Dirac insulator \citep{Yao2009Edge,gorbachev2014detecting,shimazaki2015generation},
the anomalous Hall currents from opposite valleys add up rather than
counteract in QNAHE. Therefore, it is robust against the intervalley
scattering and can be used to probe the quantized valley Chern number.
As for the inverse Faraday effect of Dirac materials, it has been
proposed through the second-order response that
contains both intraband and interband contributions in doped semimetals \citep{Tokman2020Inverse2,gao2020topological,Liang2021Axial}.
In our case, there is no free charge carrier in the Dirac insulator and
the acoustic frequency is much smaller than the band gap. Therefore,
the proposed IAFE depends only on the intrinsic band topology characterized
by the quantized valley Chern number in the presence of TRS. 
{We emphasize that both QNAHE and IAFE in the Dirac insulator are independent of the quasiparticle
lifetime due to their topological origins.}

In summary, we investigate the {quantized} nonlinear topological AE response of the Dirac insulator
that gives rise to QNAHE and IAFE. The AE conductivity and MA susceptibility are both proportional
to the quantized valley Chern number and independent of the quasiparticle lifetime. 
The rectified topological current and magnetization
generated by different acoustic waves are estimated using the realistic
parameters of monolayer hBN, which are measurable within the current experimental resolution. Owing to the large band gap of various
Dirac insulators, the proposed QNAHE and IAFE are robust and even
accessible at room temperature.

\begin{acknowledgements} \textit{Acknowledgments.}\textemdash We
thank Yu Zhang, Di Xiao, Kai Sun, Marcel Franz, Avadh Saxena, Pavlo Suckhachov, Matthias Geilhufe, and Bin Yan 
for stimulating discussions. 
The work at LANL was carried out under the auspices of the U.S. DOE NNSA under
contract No. 89233218CNA000001 through the LDRD Program, and was supported
by the Center for Nonlinear Studies at LANL, and was performed, in
part, at the Center for Integrated Nanotechnologies, an Office of
Science User Facility operated for the U.S. DOE Office of Science,
under user proposals \#2018\ensuremath{B}\ensuremath{U}0010 and \#2018\ensuremath{B}\ensuremath{U}0083. AVB is supported by Quantum CT,  European Research Council under the European Union Seventh Framework ERS-2018-SYG 810451 HERO and Knut and Alice Wallenberg Foundation KAW 2019.0068. 
\end{acknowledgements}

\bibliography{references}

\ifarXiv
    

\section*{Supplementary material}

# Supplemental Material: Quantum Nonlinear Acoustic Hall Effect and Inverse Acoustic Faraday Effect in Dirac Insulators

Ying Su,[1, *] Alexander V. Balatsky,[2, 3] and Shi-Zeng Lin[4, 1, †]

[1]*Center for Integrated Nanotechnology, Los Alamos National Laboratory, Los Alamos, New Mexico 87545, USA*
[2]*Nordita, KTH Royal Institute of Technology and Stockholm University 106 91 Stockholm, Sweden*
[3]*Department of Physics, University of Connecticut, Storrs, Connecticut 06269, USA*
[4]*Theoretical Division, T-4 and CNLS, Los Alamos National Laboratory, Los Alamos, New Mexico 87545, USA*
(Dated: July 1, 2024)

## CONTENTS

## I. SYMMETRY PROPERTIES OF THE ACOUSTOELECTRIC CONDUCTIVITY

Here we analyze the symmetry properties of the acoustoelectric (AE) conductivity with respect to the two-fold rotation $C_{2\parallel}$ and time-reversal symmetry (TRS). The two-fold rotation $C_{2\parallel}$ is along the axis of the SAW wavevector $\boldsymbol{q}$, as shown in Fig. S1. We will explain why $\sigma^{\parallel\parallel\parallel} = \sigma^{\perp\perp\perp} = \sigma^{\parallel zz} = 0$ ($\sigma^{\perp\parallel\parallel} = \sigma^{\parallel\perp\perp} = \sigma^{\perp zz} = 0$) when the SAW propagates in the zigzag (armchair) direction of the honeycomb lattice, as shown in Figs. 1 (b)-(d) of the main text. We will show how the AE conductivity transform under the TRS.

### A. Two-fold rotation symmetry $C_{2\parallel}$

Under the two-fold rotation $C_{2\parallel}$, the longitudinal current $j_\parallel$ is even (invariant), while the transverse current $j_\perp$ is odd (reverse sign). Similarly for the SAW, the longitudinal component $u_\parallel$ is even, while the perpendicular component $u_\perp$ and out-of-plane component $u_z$ are both odd under $C_{2\parallel}$. These transformations are independent on the direction of the SAW wavevector $\boldsymbol{q}$.

On the other hand, the inversion symmetry breaking makes the A and B sublattices of the honeycomb lattice inequivalent. When the SAW propagates in the zigzag direction, as shown in Fig. S1(a), $C_{2\parallel}$ exchanges the two sublattices and hence the Dirac mass $m$ (induced by the sublattice potential) is odd under the transformation. However, the Dirac mass $m$ is even under $C_{2\parallel}$ when the SAW propagates in the armchair direction, as shown in Fig. S1(b). In Table S1, we summarize how $m$, $u_\parallel$, $u_\perp$, $u_z$, $j_\parallel$, and $j_\perp$ transform under the transformation $C_{2\parallel}$ when the SAW propagates in the zigzag and armchair directions.

According to Eqs. (6) and (7) of the main text, the AE conductivity is proportional to the valley Chern number $\mathrm{sgn}(m)$ and depends on the polar angle $\theta$ of $\boldsymbol{q}$. In general, we can express the AE conductivities of linear and nonlinear currents as

$$\sigma^{\alpha\beta} = \mathrm{sgn}(m)f^{\alpha\beta}(\theta), \qquad \sigma^{\alpha\beta\gamma} = \mathrm{sgn}(m)f^{\alpha\beta\gamma}(\theta), \tag{S1}$$

and then Eq. (8) of the main text becomes

$$j_\alpha^{(1)} = \mathrm{sgn}(m)f^{\alpha\beta}(\theta)u_\beta, \qquad j_\alpha^{(2)} = \mathrm{sgn}(m)f^{\alpha\beta\gamma}(\theta)u_\beta u_\gamma. \tag{S2}$$

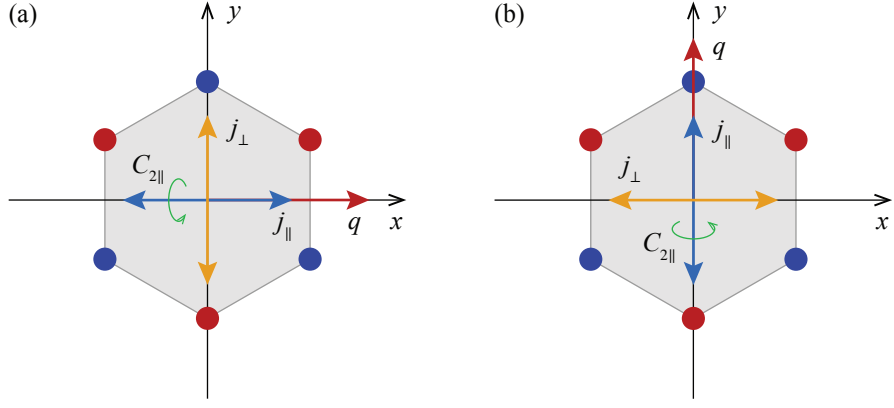

Figure S1. Schematic diagrams of the $C_{2\parallel}$ rotation along the axis of the SAW wavevector $\boldsymbol{q}$ that is the zigzag (a) and armchair (b) directions of the honeycomb lattice.

| direction of $\boldsymbol{q}$ | $m$ | $u_\parallel$ | $u_\perp$ | $u_z$ | $j_\parallel$ | $j_\perp$ |
|---|---|---|---|---|---|---|
| zigzag | odd | even | odd | odd | even | odd |
| armchair | even | even | odd | odd | even | odd |

Table S1. Transformation of $m$, $u_\parallel$, $u_\perp$, $u_z$, $j_\parallel$, and $j_\perp$ under the two-fold rotation $C_{2\parallel}$ along the axis of the SAW wavevector $\boldsymbol{q}$, when the SAW vector $\boldsymbol{q}$ is in the zigzag or armchair direction.

Without specifying the form of $f^{\alpha\beta}(\theta)$ and $f^{\alpha\beta\gamma}(\theta)$, we can deduce when they must vanish from the symmetry constraint. First of all, the two sides of Eq. (S2) have to transform in the same manner under the symmetry operation. On the left-hand side, we have $j_\parallel$ even and $j_\perp$ odd under $C_{2\parallel}$. On the right-hand side, when $m$ is odd for $\boldsymbol{q}$ in the zigzag direction with $\theta = n\pi/3$, $\mathrm{sgn}(m)u_\parallel$ and $\mathrm{sgn}(m)u_z^2$ are both odd while $\mathrm{sgn}(m)u_\perp$ is even. Therefore, $f^{\parallel\parallel}(n\pi/3) = f^{\perp\perp}(n\pi/3) = f^{\parallel zz}(n\pi/3) = 0$ because the two sides of Eq. (S2) do not match under $C_{2\parallel}$. In the same way, we can infer $f^{\perp\parallel}(n\pi/3 + \pi/6) = f^{\parallel\parallel}(n\pi/3 + \pi/6) = f^{\perp zz}(n\pi/3 + \pi/6) = 0$ when $\boldsymbol{q}$ is in the armchair direction with $\theta = n\pi/3 + \pi/6$. From this symmetry analysis, we explain why $\sigma^{\parallel\parallel} = \sigma^{\perp\perp} = \sigma^{\parallel zz} = 0$ ($\sigma^{\perp\parallel} = \sigma^{\parallel\perp} = \sigma^{\perp zz} = 0$) when the SAW propagates in the zigzag (armchair) direction.

### B. Time-reversal symmetry

Under the TRS, the topological currents $j_\alpha^{(n)}$ (where $\alpha = \parallel$ or $\perp$ and $n = 1$ or $2$) in Eqs. (6) and (7) of the main text are odd, while the time-independent SAW amplitudes $u_\beta$ (where $\beta = \parallel$, $\perp$, and $z$) in Eq. (5) are even. Since they always transform oppositely under the TRS, the AE conductivity tensors $\sigma_{\alpha\beta}$ and $\sigma_{\alpha\beta\gamma}$ must be TRS odd such that the two sides of Eq. (8) transform equally under the TRS. This can be easily verified from their expressions

$$\begin{pmatrix} \sigma_{\parallel\parallel} \\ \sigma_{\perp\parallel} \end{pmatrix} = \frac{2e^2\mathrm{sgn}(m)gq\omega}{h}\begin{pmatrix} \sin 3\theta \\ \cos 3\theta \end{pmatrix}, \qquad \begin{pmatrix} \sigma_{\parallel\perp} \\ \sigma_{\perp\perp} \end{pmatrix} = \frac{2e^2\mathrm{sgn}(m)gq\omega}{h}\begin{pmatrix} \cos 3\theta \\ -\sin 3\theta \end{pmatrix}, \tag{S3}$$

$$\begin{pmatrix} \sigma_{\parallel zz} \\ \sigma_{\perp zz} \end{pmatrix} = -i\frac{e^2\mathrm{sgn}(m)gq^2\omega}{h}\begin{pmatrix} \sin 3\theta \\ \cos 3\theta \end{pmatrix}, \tag{S4}$$

that are obtained from Eqs. (6)-(8) of the main text. Here we have the valley Chern number $\mathrm{sgn}(m)$, angular frequency $\omega$, and wave vector $q$ being TRS odd, while the other physical quantities are TRS even. Note that the imaginary number is also TRS odd. Therefore, the AE conductivities in Eqs. (S3) and (S4) are TRS odd. In Table S2, we list how these physical quantities transform under the TRS.

| $j_\alpha^{(n)}$ | $u_\beta$ | $\sigma_{\alpha\beta}$ | $\sigma_{\alpha\beta\gamma}$ | sgn($m$) | $\omega$ | $q$ |
|---|---|---|---|---|---|---|
| odd | even | odd | odd | odd | odd | odd |

Table S2. Transformation of $j_\alpha^{(n)}$, $u_\beta$, $\sigma_{\alpha\beta}$, $\sigma_{\alpha\beta\gamma}$, sgn($m$), $\omega$, and $q$ under the TRS.

## II. STRAIN-INDUCED CHARGE DENSITY

To verify the topological AE response, we compare the strain-induced charge density from Eq. (4) of the main text with the one calculated from the tight-binding model. Here we consider a honeycomb lattice described by

$$H = \sum_i \varepsilon_i c_i^\dagger c_i + \sum_{\langle ij \rangle} t_{ij} c_i^\dagger c_j, \tag{S5}$$

where $\varepsilon_i = \pm m$ is the sublattice potential and $t_{ij}$ is the hopping energy between two nearest neighboring sites at $\boldsymbol{r}_i$ and $\boldsymbol{r}_j$. Under strain, the lattice bond length $d_{ij}$ is altered as

$$d_{ij} = \sqrt{\boldsymbol{d}_{ij}^2} = \sqrt{a_0^2 + 2\boldsymbol{\delta}_{ij} \cdot \overleftrightarrow{\boldsymbol{u}} \cdot \boldsymbol{\delta}_{ij} + (\overleftrightarrow{\boldsymbol{u}} \cdot \boldsymbol{\delta}_{ij})^2}, \tag{S6}$$

where $\boldsymbol{d}_{ij} = \boldsymbol{r}_i + \boldsymbol{u}(\boldsymbol{r}_i) - \boldsymbol{r}_j - \boldsymbol{u}(\boldsymbol{r}_j) = \boldsymbol{\delta}_{ij} + \overleftrightarrow{\boldsymbol{u}} \cdot \boldsymbol{\delta}_{ij}$ with the unstrained bond vector $\boldsymbol{\delta}_{ij} = \boldsymbol{r}_i - \boldsymbol{r}_j$ whose length $|\boldsymbol{\delta}_{ij}| = a_0$ is constant. $\overleftrightarrow{\boldsymbol{u}}$ denotes the strain tensor and the hopping energy is

$$t_{ij} = -t_0 \exp\left[-\beta\left(\frac{d_{ij}}{a_0} - 1\right)\right], \tag{S7}$$

where $\beta$ is the Grüneisen parameter characterizing the decay rate of the hopping energy with respect to the bond length. By Fourier transform into the momentum space and then expanding the Hamiltonian around the $K$ and $K'$ valleys to the linear order in terms of momentum and strain tensor, the effective Dirac action in Eq. (1) of the main text can be derived [S1].

Without loss of generality, we take the monolayer hBN as an example and calculate the strain-induced charge density at half-filling from the tight-binding model. For monolayer hBN, the model parameters are $\beta = 3.3$, $a_0 = 1.44$ Å , $t_0 = 2.16$ eV, and $m = 3$ eV [S2]. Two types of strain, that are induced by SAW and triaxial stretch in the manuscript, are considered here. Furthermore, we will discuss how the bulk and edge charges vary with the system size.

### A. Surface acoustic wave

For the SAW in Eq. (5) of the main text, the lattice displacement at time $t = 0$ s is displayed in Fig. S2(a). Here we set the SAW amplitudes $u_\parallel = 0.2a_0$, $u_\perp = 0.3a_0$, $u_z = 5a_0$, wavelength $2\pi/q = 100a_0$, and polar angle of the wave vector $\theta = 80°$. The in-plane displacement is enlarged by 10 times to demonstrate the lattice deformation in Fig. S2(a). According to the continuum theory in the main text, the SAW-induced pseudo-magnetic field $B_s$ and charge density $\rho$ are shown in Figs. S2(b) and (c), respectively. They are proportional to each other but with an opposite sign, as shown in Eq. (4) of the main text.

In Fig. S2(d), we show the SAW-induced charge density from the tight-binding model Eq. (S5). Here we sum over the charge density of all eigenstates below the band gap of the monolayer hBN, as shown in Fig. S3. The uniform charge density from ions is also considered here such that the whole system is charge neutral at half filling. The color of each hexagon in Fig. S2(d) represents the charge density within each unit cell. Figs. S2(c) and (d) exhibit excellent agreement with each other except that the charge density amplitude in Fig. S2(c) is about 5 times larger than that in Fig. S2(d). This is because only the Dirac quasiparticle is considered in the continuum theory, however, there are also high-energy states in the lattice model. These states come from the higher order terms of the Hamiltonian away from the $K$ and $K'$ valleys and their response to the strain does not follow Eq. (4) of the main text.

### B. Triaxial stretch

For the triaxial stretch

$$(u_r, u_\theta) = \frac{r^2}{u_0}(\cos(3\phi - 3\theta), \sin(3\phi - 3\theta)), \tag{S8}$$

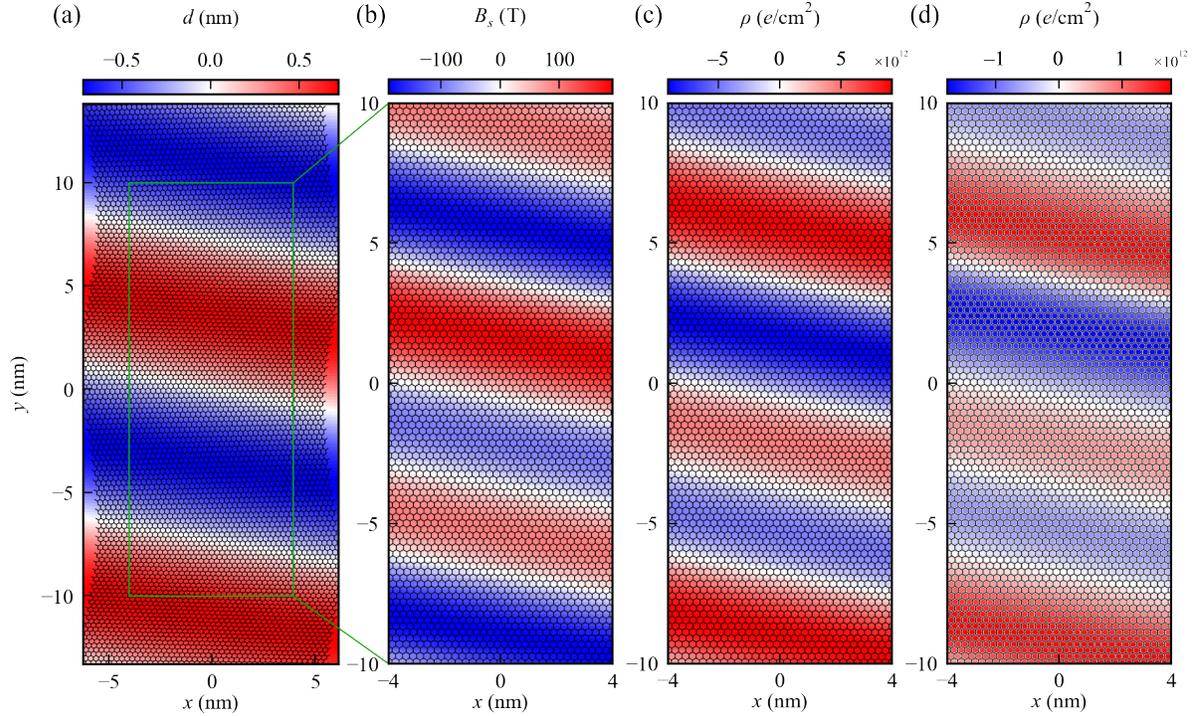

Figure S2. (a) Lattice deformation of a monolayer hBN subject to a SAW. Here we enlarge the in-plane displacement by 10 times to visualize the lattice deformation. (b) and (c) The SAW-induced pseudo-magnetic field $B_s$ and charge density $\rho$ in the bulk of the monolayer hBN in (a). These results are obtained from the quantum field theory. (d) The charge density calculated from the tight-bind model and corresponds to that in (c).

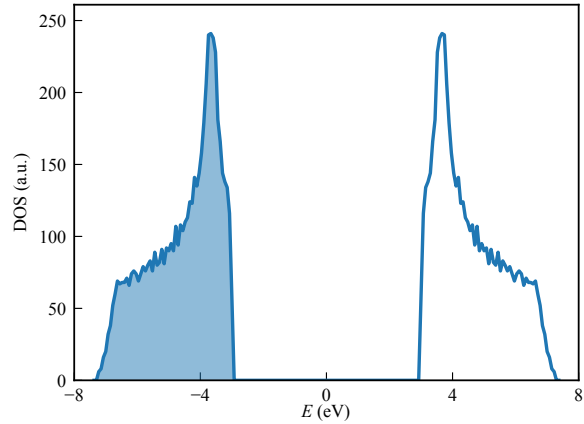

Figure S3. Density of states of the monolayer hBN in Fig. S2(a). At half filling, all states below the band gap are fully filled.

considered in the main text, a uniform charge density

$$\rho = -\frac{16e^2\mathrm{sgn}(m)g}{hu_0}\sin 3\phi, \tag{S9}$$

can be generated in the bulk. Here we verify this result by simulating the charge density distribution of a monolayer hBN under the triaxial stretch from the tight-binding model Eq. S5. For a monolayer hBN flake, as shown in Fig. S4(a), there are alternating positive and negative charge accumulation at the adjacent edges, that stems from the edge states of zigzag edges with different sublattices [S3]. Note that these edge states are not topologically protected and the edge charge distribution depends on the shape of edges. Furthermore, there is no net charge in the bulk, as shown in the low panel of Fig. S4.

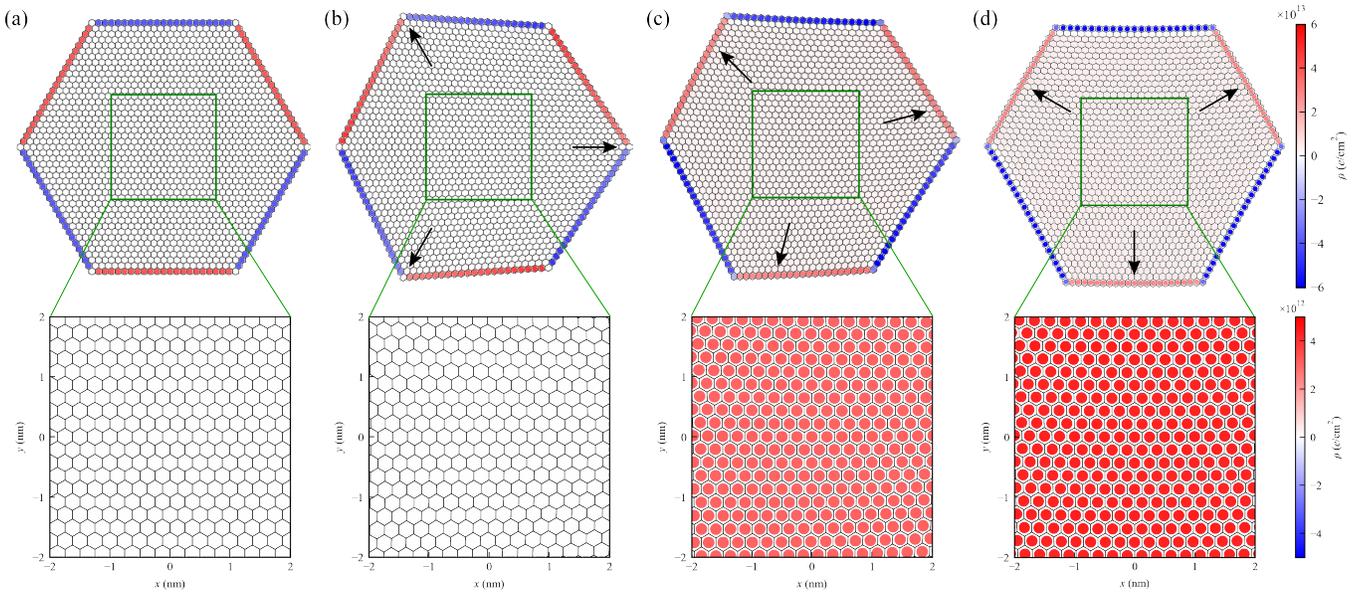

Figure S4. (a) Distribution of charge density in a hBN flake without strain. The lower panel zooms into the bulk of the upper panel. (b)-(d) The charge density in the hBN flakes subject to the triaxial stretch with the polar angle $\phi = 0°$, $15°$, and $30°$, respectively. The black arrows represent the directions of the stretch.

Now we apply the triaxial stretch to the monolayer hBN for different polar angle $\phi$ of the stretch axis. Here we set $u_0 = 1000 a_0$ that controls the strength of the strain. In Figs. S4(b)-(d), we show the charge density distribution of the hBN flake with $\phi = 0°$, $15°$, and $30°$, respectively. There is no net bulk charge for $\phi = 0°$, as predicted by Eq. (S9). As $\phi$ ramps up, the uniform bulk charge density $\rho$ gradually increases and reaches its maximum at $\phi = 30°$. The variation of $\rho$ with respect to the polar angle $\phi$ has a period of $2\pi/3$, as demonstrated in Eq. (S9), that is consistent with the $C_{3z}$ symmetry of the hBN lattice.

### C. Scaling of bulk and edge charges

Since the hBN flake is charge neutral at half filling, the net charge in the bulk indicates that there must be equal amount of opposite charge at the edges. For the triaxial stretch considered above, the bulk charge density is uniform and hence the total bulk charge $Q_{\text{bulk}}$ scales linearly with the area of the system as

$$Q_{\text{bulk}} \propto \rho L^2, \tag{S10}$$

where $L$ denotes the side length of the system. Then the total edge charge with an opposite sign is

$$Q_{\text{edge}} = -Q_{\text{bulk}} \propto \overline{\rho}_{\text{edge}} L, \tag{S11}$$

such that the mean edge charge density becomes

$$\overline{\rho}_{\text{edge}} \propto -\rho L. \tag{S12}$$

This peculiar behavior indicates that the mean edge charge density increases linearly with edge length of the system.

## III. DERIVATION OF THE MAGNETOACOUSTIC SUSCEPTIBILITY

To derive the magnetoacoustic (MA) susceptibility in Eq. (13) of the main text, we first need to solve the displacement field by minimizing the elastic energy functional in Eq. (11) that yields

$$\lambda \partial_x (u_{xx} + u_{yy}) + 2\mu (\partial_x u_{xx} + \partial_y u_{xy}) = 0, \tag{S13}$$

$$\lambda \partial_y (u_{xx} + u_{yy}) + 2\mu (\partial_y u_{yy} + \partial_x u_{xy}) = 0. \tag{S14}$$

After the Fourier transformation, the strain tensor becomes

$$u_{ij}(\boldsymbol{r}) = \sum_{\boldsymbol{k}} u_{\boldsymbol{k},ij} e^{i\boldsymbol{k}\cdot\boldsymbol{r}} = \sum_{\boldsymbol{k}} \frac{1}{2} \left[ ik_i u_{\boldsymbol{k},j} + ik_j u_{\boldsymbol{k},i} + f_{\boldsymbol{k},ij} \right] e^{i\boldsymbol{k}\cdot\boldsymbol{r}}, \tag{S15}$$

where

$$u_i(\boldsymbol{r}) = \sum_{\boldsymbol{k}} u_{\boldsymbol{k},i} e^{i\boldsymbol{k}\cdot\boldsymbol{r}}, \quad d(\boldsymbol{r}) = \sum_{\boldsymbol{k}} d_{\boldsymbol{k}} e^{i\boldsymbol{k}\cdot\boldsymbol{r}}, \quad \partial_i d(\boldsymbol{r}) \partial_j d(\boldsymbol{r}) = \sum_{\boldsymbol{k}} f_{\boldsymbol{k},ij} e^{i\boldsymbol{k}\cdot\boldsymbol{r}}, \quad f_{\boldsymbol{k},ij} = -\sum_{\boldsymbol{k}'} k_i' \left( k_j - k_j' \right) d_{\boldsymbol{k}'} d_{\boldsymbol{k}-\boldsymbol{k}'}. \tag{S16}$$

Then Eqs. (S13) and (S14) yield

$$(\lambda + 2\mu) k_x u_{\boldsymbol{k},xx} + \lambda k_x u_{\boldsymbol{k},yy} + 2\mu k_y u_{\boldsymbol{k},xy} = 0, \tag{S17}$$

$$(\lambda + 2\mu) k_y u_{\boldsymbol{k},yy} + \lambda k_y u_{\boldsymbol{k},xx} + 2\mu k_x u_{\boldsymbol{k},xy} = 0, \tag{S18}$$

that can be solved analytically. This gives the displacement field

$$u_{\boldsymbol{k},x} = i \frac{f_{\boldsymbol{k},xy} k_x \left[ (\lambda + 2\mu) k_x^2 + (3\lambda + 4\mu) k_y^2 \right] + \left[ \lambda k_x^2 - (\lambda + 2\mu) k_y^2 \right] \left( f_{\boldsymbol{k},yy} k_x - 2 f_{\boldsymbol{k},xy} k_y \right)}{2 (\lambda + 2\mu) \left( k_x^2 + k_y^2 \right)^2}, \tag{S19}$$

$$u_{\boldsymbol{k},y} = i \frac{k_x^2 k_y \left[ (3\lambda + 4\mu) f_{\boldsymbol{k},yy} - (\lambda + 2\mu) f_{\boldsymbol{k},xx} \right] + k_y^3 \left[ \lambda f_{\boldsymbol{k},xx} + (\lambda + 2\mu) f_{\boldsymbol{k},yy} \right] + 2 f_{\boldsymbol{k},xy} k_x \left[ (\lambda + 2\mu) k_x^2 - \lambda k_y^2 \right]}{2 (\lambda + 2\mu) \left( k_x^2 + k_y^2 \right)^2}. \tag{S20}$$

From $\boldsymbol{u}_{\boldsymbol{k}}$, we can obtain the emergent gauge field

$$\boldsymbol{a}(\boldsymbol{r}) = \sum_{\boldsymbol{k}} \boldsymbol{a}_{\boldsymbol{k}} e^{i\boldsymbol{k}\cdot\boldsymbol{r}} = g \sum_{\boldsymbol{k}} \begin{pmatrix} u_{\boldsymbol{k},yy} - u_{\boldsymbol{k},xx} \\ 2 u_{\boldsymbol{k},xy} \end{pmatrix} e^{i\boldsymbol{k}\cdot\boldsymbol{r}}, \tag{S21}$$

$$\boldsymbol{a}_{\boldsymbol{k}} = g \begin{pmatrix} u_{\boldsymbol{k},yy} - u_{\boldsymbol{k},xx} \\ 2 u_{\boldsymbol{k},xy} \end{pmatrix} = g \begin{pmatrix} \frac{(\lambda+\mu)(k_x^2 - k_y^2)\left[ k_x^2 f_{\boldsymbol{k},yy} - 2k_x k_y f_{\boldsymbol{k},xy} + k_y^2 f_{\boldsymbol{k},xx} \right]}{(\lambda+2\mu)(k_x^2 + k_y^2)^2} \\ -\frac{2(\lambda+\mu) k_x k_y \left[ k_x^2 f_{\boldsymbol{k},yy} - 2k_x k_y f_{\boldsymbol{k},xy} + k_y^2 f_{\boldsymbol{k},xx} \right]}{(\lambda+2\mu)(k_x^2 + k_y^2)^2} \end{pmatrix}, \tag{S22}$$

and the pseudo-magnetic field [S4]

$$B_s(\boldsymbol{r}) = \sum_{\boldsymbol{k}} B_{s,\boldsymbol{k}} e^{i\boldsymbol{k}\cdot\boldsymbol{r}} = (\nabla \times \boldsymbol{a}) \cdot \hat{\boldsymbol{z}}, \tag{S23}$$

$$B_{s,\boldsymbol{k}} = g \left[ -2ik_x u_{\boldsymbol{k},xy} - ik_y \left( u_{\boldsymbol{k},xx} - u_{\boldsymbol{k},yy} \right) \right] = i \frac{g(\lambda+\mu) k_y \left( 3k_x^2 - k_y^2 \right) \left[ k_x^2 f_{\boldsymbol{k},yy} - 2k_x k_y f_{\boldsymbol{k},xy} + k_y^2 f_{\boldsymbol{k},xx} \right]}{(\lambda+2\mu) \left( k_x^2 + k_y^2 \right)^2}. \tag{S24}$$

According to Eq. (4) of the main text, the strain-induced charge density is proportional to the pseudo-magnetic field as

$$\rho(\boldsymbol{r}) = \sum_{\boldsymbol{k}} \rho_{\boldsymbol{k}} e^{i\boldsymbol{k}\cdot\boldsymbol{r}} = -\frac{2e^2 \text{sgn}(m)}{h} B_s, \tag{S25}$$

$$\rho_{\boldsymbol{k}} = -\frac{2e^2 \text{sgn}(m)}{h} B_{s,\boldsymbol{k}} = -i \frac{2e^2 \text{sgn}(m) g(\lambda+\mu) k_y \left( 3k_x^2 - k_y^2 \right) \left[ k_x^2 f_{\boldsymbol{k},yy} - 2k_x k_y f_{\boldsymbol{k},xy} + k_y^2 f_{\boldsymbol{k},xx} \right]}{h(\lambda+2\mu) \left( k_x^2 + k_y^2 \right)^2}. \tag{S26}$$

Then the electric charge can be driven into orbital motion by a circularly polarized acoustic wave that gives rise to a static magnetization, i.e., the inverse acoustic Faraday effect (IAFE) proposed in the manuscript. The static magnetization is

$$\boldsymbol{M}(\boldsymbol{r}) = \frac{1}{2} \rho(\boldsymbol{r}) \boldsymbol{u}' \times \dot{\boldsymbol{u}}', \tag{S27}$$

where $\boldsymbol{u}' = \text{Re}[u'(\hat{\boldsymbol{x}} - i\hat{\boldsymbol{y}}) e^{i(q_z z - \omega t)}]$ is the circularly polarized acoustic wave propagating in the out-of-plane direction. After the Fourier transformation, Eq. (S27) yields

$$\boldsymbol{M}_{\boldsymbol{k}} = \chi_{\boldsymbol{k}} \boldsymbol{u}'_{q_z,\omega} \times \boldsymbol{u}'^{*}_{q_z,\omega}, \tag{S28}$$

where the MA susceptibility is

$$\chi_{\boldsymbol{k}} = i\omega \rho_{\boldsymbol{k}} = \frac{2e^2 \text{sgn}(m) g\omega(\lambda+\mu) k_y \left( 3k_x^2 - k_y^2 \right) \left[ k_x^2 f_{\boldsymbol{k},yy} - 2k_x k_y f_{\boldsymbol{k},xy} + k_y^2 f_{\boldsymbol{k},xx} \right]}{h(\lambda+2\mu) \left( k_x^2 + k_y^2 \right)^2}, \tag{S29}$$

as written in Eq. (13) of the main text, and $\boldsymbol{u}'(\boldsymbol{r}) = \sum_{\boldsymbol{k}} u'_{\boldsymbol{k},\omega} e^{i(\boldsymbol{k}\cdot\boldsymbol{r} - \omega t)}$ with $\boldsymbol{u}'_{q_z,\omega} = (u'/2, -iu'/2)$.

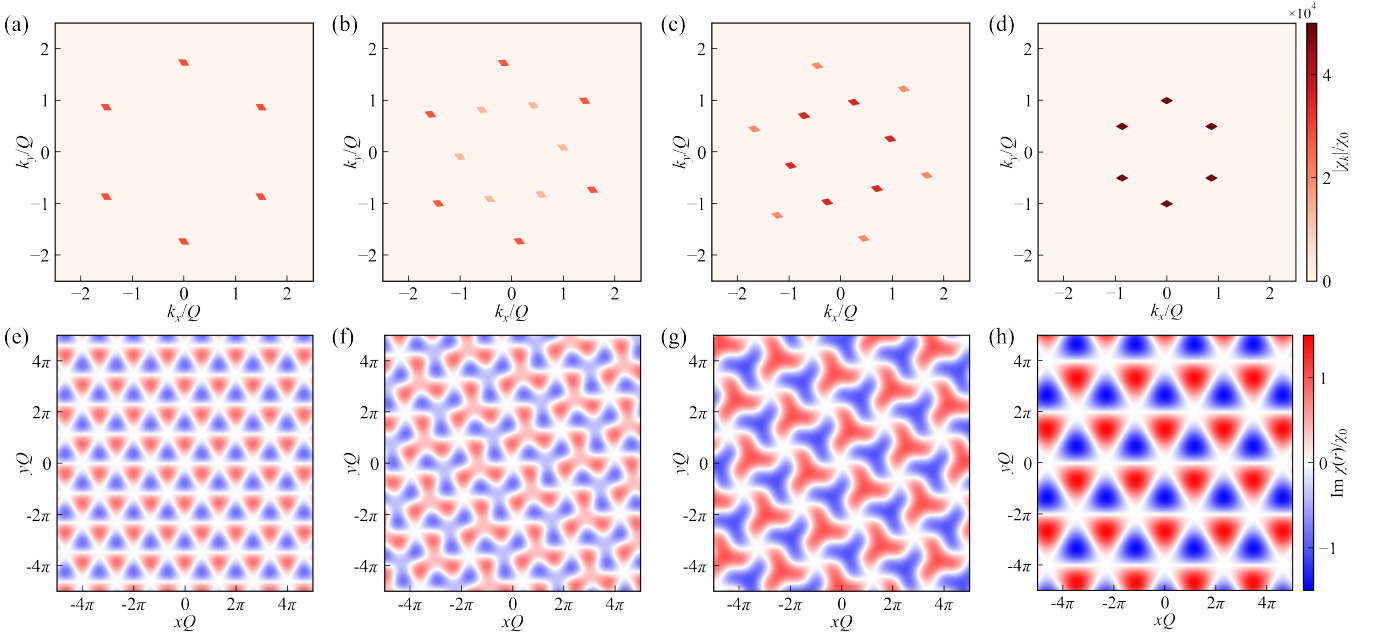

Figure S5. (a)-(d) Normalized MA susceptibilities of the periodically rippled Dirac insulator considered in the main text and with the polar angles $\theta = 0°$, $5°$, $15°$, and $30°$, respectively. The normalization factor $\chi_0$ is defined in Eq. (14) of the main text. (e)-(h) The real-space distribution of the normalized MA susceptibility, corresponding to that in (a)-(d).

## IV. POLAR ANGLE DEPENDENCE OF THE MAGNETOACOUSTIC SUSCEPTIBILITY

As demonstrated in Fig. 2(b) and (d) of the main text, the MA susceptibility $\chi_{\boldsymbol{k}}$ peaks at $\pm \boldsymbol{Q}_j$ and $\pm(\boldsymbol{Q}_{j+1} - \boldsymbol{Q}_j)$ exhibiting different polar angle dependences. In particular, for $\boldsymbol{Q}_j$ along the zigzag (armchair) direction with the polar angle $\theta = n\pi/3$ ($n\pi/3 + \pi/6$), $|\chi_{\boldsymbol{Q}_1}|$ ($|\chi_{\boldsymbol{Q}_1 - \boldsymbol{Q}_3}|$) vanishes that leads to the magnetization with different periods. Here we show this results explicitly by plotting the MA susceptibility in both momentum and real spaces for different polar angles $\theta = 0°$, $5°$, $15°$, $30°$, as displayed in Figs. S5. As $\theta$ increases from $0°$ to $30°$, the MA susceptibility initially exhibit six peaks at $\pm(\boldsymbol{Q}_{j+1} - \boldsymbol{Q}_j)$ that gradually decreases and eventually vanishes, as shown in Figs. S5(a)-(d). Meanwhile, the peaks at $\pm \boldsymbol{Q}_j$ ramps up and reaches their maximum until $\theta = 30°$. In real space, the corresponding $\chi(\boldsymbol{r})$ are shown in Figs. S5(e)-(h). Interestingly, $\chi(\boldsymbol{r})$ for $\theta = 0°$ and $30°$ in Figs. S5(e) and (h) exhibit similar patterns but their periods differ by $\sqrt{3}$ times.

* yingsu001@gmail.com
† szl@lanl.gov


\fi

\end{document}